\documentclass[conference,compsoc]{IEEEtran}
\usepackage{cite}
\usepackage{amsmath,amssymb,amsfonts}
\usepackage{graphicx}
\usepackage{textcomp}
\usepackage{xcolor}
\usepackage{hyperref}
\usepackage{bm}
\usepackage{multirow}
\usepackage{algorithm}  
\usepackage{algpseudocode}  
\usepackage[misc,geometry]{ifsym}
\def\BibTeX{{\rm B\kern-.05em{\sc i\kern-.025em b}\kern-.08em
    T\kern-.1667em\lower.7ex\hbox{E}\kern-.125emX}}

\usepackage[colorinlistoftodos]{todonotes}

\begin{document}

\title{East: Efficient and Accurate Secure Transformer Framework for Inference\\
\thanks{Identify applicable funding agency here. If none, delete this.}
}

\author{Yuanchao Ding$^{1,2}$, Hua Guo$^{1,2,\textrm{\Letter}}$, Yewei Guan$^{1}$, Weixin Liu$^{1}$, Jiarong Huo$^{1}$, Zhenyu Guan$^{1}$,  Xiyong Zhang$^{3}$
\\
\IEEEauthorblockA{
1. School of Cyber Science and Technology, Beihang University, Beijing, China, \\ 
\{dych21,hguo,ame\_reiori,19373768,20373239,guanzhenyu\}@buaa.edu.cn
  \\
2. State Key Laboratory of Software Development Environment, Beihang University, Beijing, China,
   \\
3. Beijing Institute of Satellite Information Engineering, Beijing, China, \\ Xiyong.Zhang@hotmail.com
}}

\maketitle

\begin{abstract}
Transformer has been successfully used in practical applications, such as ChatGPT, due to its powerful advantages.  However, users' input is leaked to the model provider during the service. With people's attention to privacy, privacy-preserving Transformer inference is on the demand of such services.  Secure protocols for non-linear functions are crucial in privacy-preserving Transformer inference, which are not well studied. Thus, designing
practical secure protocols for non-linear functions is hard but significant to model performance. In this work, we propose a framework \emph{East} to enable efficient and accurate secure Transformer inference.   
Firstly, we propose a new oblivious piecewise polynomial evaluation algorithm and apply it to the activation functions, which reduces the runtime and communication of GELU by over 1.5$\times$ and 2.5$\times$, compared to prior arts. Secondly, the secure protocols for softmax and layer normalization are carefully designed to faithfully maintain the desired functionality. Thirdly, several optimizations are conducted in detail to enhance the overall efficiency. 
We applied \emph{East} to BERT and the results show that the inference accuracy remains consistent with the plaintext inference without fine-tuning. Compared to Iron, we achieve about 1.8$\times$ lower communication within 1.2$\times$ lower runtime.
\end{abstract}


\section{Introduction}
With the development of artificial intelligence, Transformer has shown good performance in the fields of natural language processing (NLP) and computer vision (CV) since its birth\cite{vaswani2017attention}. Recently, \cite{Openai} showed the success of emerging inference services and applications of Transformer.
These services typically deploy trained models to the cloud, where the main challenge is to ensure the privacy of client's input data and server's model parameters when inferring, which is called privacy-preserving Transformer inference (PPTi). 

Secure multi-party computation (MPC), providing stronger security and similar accuracy to plaintext models, is a widely used technique in privacy-preserving machine learning (PPML) and has attracted much attention.
Before the exploration of PPTi based on MPC, a number of works on secure two-party neural network inference (2PC-NNi) have been proposed since the proposal of SecureML \cite{mohassel2017secureml}, such as  MiniONN \cite{liu2017oblivious},  Chameleon \cite{riazi2018chameleon}, Gazelle \cite{juvekar2018gazelle} and Delphi \cite{mishra2020delphi}. These works use homomorphic encryption (HE), oblivious transfer (OT), garbled circuit (GC), and secret sharing (SS) to conduct 2PC-NNi on several small neural networks. CrypTFlow2 \cite{CrypTFlow2} resolves the truncation issue and builds a comparison protocol based on SS and OT, which can be used to construct ReLU and obtain better results than previous works using GC. Thus, CrypTFlow2 can execute inference over larger networks, e.g., ImageNet-scale DNNs. Another excellent work, Cheetah \cite{Cheetah}, designs efficient protocols for linear layers based on HE and provides a fast truncation protocol that allows for one-bit error, as well as a faster implementation with known the most significant bit (MSB). While 2PC-NNi works can contribute to PPTi, there are still some key aspects that need to be considered due to different model architectures, especially some non-linear functions, such as GELU\cite{GELU}, tanh, softmax, and layer normalization (LN).
\par
THE-X \cite{THE-X} is the first work to explore PPTi using cryptographic techniques. In THE-X \cite{THE-X}, GELU is replaced by ReLU, and tanh is unsupported. THE-X \cite{THE-X} uses approximation methods for softmax and LN to convert Transformer into the function supported by fully HE operations. There are several issues with THE-X \cite{THE-X}. (1) THE-X \cite{THE-X} lacks protection for the privacy of the server because THE-X \cite{THE-X} leaks the intermediate results to the client when computing ReLU, which may cause server privacy leakages. (2) The average decrease of THE-X on accuracy \cite{THE-X} is over 1\%  compared to the plaintext inference. (3) The original model needs to be converted into an approximate form when using THE-X \cite{THE-X}, which makes the original model parameters unusable and requires extra model retraining. Several subsequent works (\cite{li2022mpcformer}\cite{zeng2022mpcvit},\cite{liu2023llms},\cite{akimoto2023privformer}) address the first two issues by adopting more effective approximations and using MPC to ensure privacy. These works focus on MPC-friendly approximations of Transformer and ignore the third issue. In fact, when training a model in plaintext, the requirement of MPC-friendly Transformer may limit its flexibility. Similar work to accelerate computation by changing the model architecture has also been used in \cite{liang2023merge}, which performs privacy protection during text generation tasks. 

The recent work Iron \cite{Iron} provides a solution to all three issues.
Iron \cite{Iron} retains the original model architecture and constructs several privacy-preserving protocols for Transformer based on HE and SS. In Iron \cite{Iron}, the matrix multiplication is built by optimizing HE-based one from Cheetah \cite{Cheetah}, while the secure protocols for non-linear functions are devised by primitives based on look-up tables (LUT) and SS from SIRNN \cite{rathee2021sirnn}. More recently, \cite{zheng2023primer} directly uses  GC to handle non-linear functions, which is different from Iron \cite{Iron}. Another work \cite{dong2023puma} designs high-quality approximations for expensive functions and obtains better results than \cite{li2022mpcformer}, but \cite{dong2023puma} uses the technology of secure three-party computation (3PC), which works in a different scenario from 2PC. 
\par
Among the above works, to our knowledge, Iron\cite{Iron} is state-of-the-art 2PC work for PPTi. The obstacle of Iron\cite{Iron} is that the non-linear functions based on LUT bring considerable overheads. In Iron\cite{Iron}, the runtime and communication of non-linear functions account for about 80\% and 87\% of the total, respectively. Therefore, designing secure and efficient protocols for non-linear functions with high accuracy while maintaining model architecture is a challenge. 


\subsection{Our contribution}
In this paper, we design \textit{East}, an efficient and accurate secure Transformer framework for inference. The contributions of this paper are as follows:

\begin{itemize}
    \item Proposing communication-efficient protocols for the activation functions, such as GELU and tanh, through our new  oblivious evaluation of approximating functions. Our protocols achieve about 1.5$\times$ and 2.6$\times$ improvement in runtime and communication, compared to  NFGen \cite{NFGen}.
    \item Designing the protocols for softmax and LN carefully using our 
 proposed conversion methods and error-limited determination methods. Our protocols achieve about 1.3$\times$ lower communication for softmax and 1.2$\times$ lower runtime for LN, compared with Iron\cite{Iron}.
    \item Conducting several optimizations to enhance the overall performance, including combination of linear layers, truncation optimization, processing of packed communication, and handling of padding and masking.
\end{itemize}
\par
We do BERT \cite{bert} inference with \emph{East} and the results show that the inference accuracy remains consistent with the plaintext inference. Compared to Iron \cite{Iron}, we achieve about 1.2$\times$ lower runtime and 1.8$\times$ lower communication for the non-linear functions.
\par

\subsection{Our Techniques}

\textbf{Activation functions}. The previous works usually use ReLU as activation function and design privacy-preserving protocol for ReLU.  Iron \cite{Iron} designs non-linear protocols by improving some basic protocols based on LUT from SIRNN \cite{rathee2021sirnn}. Another way to solve complex non-linear functions is polynomial approximation (SecureML\cite{mohassel2017secureml}, MiniONN\cite{liu2017oblivious}), and the trade-off between the accuracy and efficiency is worth considering. Recently, NFGen \cite{NFGen} provides a paradigm for dealing with non-linear functions: determining the parameters of the polynomials in plaintext and then performing oblivious piecewise polynomial evaluation (OPPE) in secret. However, there are some issues when introducing the OPPE algorithm of NFGen \cite{NFGen} to 2PC work. NFGen \cite{NFGen} performs subtraction and multiplication on the comparison results, which are usually boolean shares (BShare), while subtraction and multiplication need arithmetic shares (AShare), which usually requires additional conversion from BShare to AShare (B2A). And multiple multiplications are used in NFGen \cite{NFGen} to evaluate polynomials, which requires more communication. In addition, the functions evaluated in NFGen \cite{NFGen} are limited to a finite interval. 
\par
To solve these problems, we propose a new OPPE without using B2A and apply a multiplexer (MUX) protocol to replace the multiplications. Overall, our new OPPE optimizes the runtime and the communication by about 1.5$\times$ and 2.6$\times$ compared to NFGen \cite{NFGen}. We also achieve about 1.5$\times$ lower runtime and 2.9$\times$ lower communication for GELU compared to Iron \cite{Iron}.
\par
\textbf{Softmax and LN}. The use of softmax in Transformer is different from DNNs. In DNNs, softmax is usually used at the last layer and the outputs of softmax represent the relative probability between different classes. The class with the highest relative probability is usually used as the classification result. Softmax does not change the classification results, so most 2PC-NNi works evaluate the networks without softmax, such as \cite{liu2017oblivious}, \cite{riazi2018chameleon},  \cite{juvekar2018gazelle}, \cite{mishra2020delphi}, \cite{CrypTFlow2} and \cite{Cheetah}. In Transformer, softmax is used to normalize the inner product results of queries and keys in self-attention and the output of softmax will be used for subsequent computation. Thus softmax cannot be omitted in Transformer. Some works replace softmax with a network (THE-X \cite{THE-X}) or ReLU-softmax (SecureML \cite{mohassel2017secureml}),  which are not flexible and may have an impact on the inference accuracy. Iron \cite{Iron} splits the secure evaluation of softmax to the max, exponent, and reciprocal, which are solved by LUT-based protocols from SIRNN. However, it brings communication overhead. 

LN is another layer considered differently in Transformer, corresponding to the batch normalization (BN) in 2PC-NNi works. Because the mean and variance are the saved results of the training stage, BN is converted into a linear layer in Cheetah \cite{Cheetah}. LN uses the mean and variance of the input data and cannot be converted into just linear operations. THE-X \cite{THE-X} ignores layer normalization and directly performs linear operations, which results in a significant accuracy decline. 
\par
In this paper, we propose the conversion methods for softmax and LN to make them privacy-preserving-friendly and use Newton's method to compute reciprocal and square-root reciprocal. For softmax, the input is handled by substructing the maximum to be non-positive. Then the exponent is computed by our OPPE. For LN, we use equivalence conversion to eliminate the division of input dimensions when computing the mean and variance. The remaining problem is computing reciprocal and square-root reciprocal, which can be realized by Newton's method. In Newton's method, a good initial value can reduce the number of iterations. We study the selection of the initial value and the number of iterations of the Newton's method within the given limited error and propose the algorithm for automatically determining the initial value and the number of iterations. The experimental results indicate that our softmax and LN achieve about 1.3$\times$ lower communication for softmax and  1.2$\times$ lower runtime for LN, compared with Iron \cite{Iron}.
\par
\textbf{Detailed optimizations}.
Several  optimizations are conducted in detail to enhance the overall performance. (1) Combine linear layers. We combine the result of linear operations in the attention part, including query, key, and value, and this has an improvement compared to independent computation. (2) Optimize truncation. The truncation protocol can be more efficient when MSB is known. (3) Process packed communication. The computation and communication can be processed simultaneously when performing the same computation on multi-dimensional data. (4) Handle padding and masking. We discard the method of ``padding then  masking" in the plaintext inference and only keep the corresponding computation for the actual processing length to reduce the overall costs.

\subsection{Organisation}
The rest of this paper is organized as follows.
In Section \ref{chap:2}, we introduce the preliminaries. Section \ref{chap:3} shows the workflow of \emph{East}. 
We present our supporting protocols in Section \ref{chap:4}. 
The experimental evaluation is provided in Section \ref{chap:7}. 
Section \ref{chap:8} concludes the whole paper.
\section{Preliminaries\label{chap:2}}

\subsection{System Model and Security}
We consider the similar private inference scenario with Cheetah \cite{Cheetah} and Iron\cite{Iron}, where the server holds a model $M$ with private
parameters $P$, and the client holds private input $X$. The
goal is to obtain the output of the model on the
input, i.e., $M(P, X)$, while the server learns nothing
about $X$ and the client learns nothing
about $P$. We provide security against a static semi-honest probabilistic polynomial time adversary $\mathcal{A}$ following the simulation
paradigm. For every adversary $\mathcal{A}$ in the real interaction, there is a simulator $\textsf{Sim}$ in the
ideal interaction, such that in any environment $\mathcal{A}$ cannot distinguish the view in real interactions from the one simulated by $\textsf{Sim}$ in ideal interactions.
\subsection{Cryptographic Primitives}
\textbf{Notations}. Let $[[n]]$ denote the set $\{0,\cdots,n-1\}$ for $n \in N$. $[x]$ and $\langle x \rangle$ represent the ciphertext of HE and shares of SS, respectively. Let
$\left\lceil \cdot \right\rceil$, $\left\lfloor \cdot \right\rfloor$ and $\left\lfloor \cdot \right\rceil$ denote the ceiling, flooring, and rounding function, respectively. For a signed
integer $x$, we write $x \gg f$ to denote the arithmetic right-shift
of $x$ by $f$-bit. $\lambda$ is the security parameter. Let $1\{\mathcal{L}\}$ denote the indicator function that
is 1 when $\mathcal{L}$ is true and 0 when $\mathcal{L}$ is false. For a floating-point number $a$, we denote by FXP the fixed-point number and  FXP$_{\ell,f}(a)$ the $(\ell,f)$-FXP representation of $a$ in the ring $\mathbb{Z}_{2^\ell}$.

\textbf{Secret Sharing and Beaver Triples}. We adopt the 2-out-of-2 arithmetic secret sharing (\cite{shamir1979share},\cite{cramer2015secure},\cite{demmler2015aby}) over the ring $\mathbb{Z}_{2^\ell}$. The sharing algorithm takes an $\ell$-bit value $x$ in $\mathbb{Z}_{2^\ell}$ as input, and outputs random sampled shares $\langle x\rangle_0$, $\langle x\rangle_1$ such that $x = \langle x\rangle_0 + \langle x\rangle_1$ in $\mathbb{Z}_{2^\ell}$. Beaver triples \cite{beaver1992efficient} are used to do element-wise ciphertext-ciphertext multiplication (MUL). Assume that two parties hold $\langle a\rangle,\langle b\rangle$ and $\langle c\rangle$ where $c=ab$. Given shared integers $\langle u\rangle$ and $\langle v\rangle$, $P_i$ locally computes $\langle e\rangle_i=\langle u\rangle_i-\langle a\rangle_i$ and $\langle f\rangle_i=\langle v\rangle_i-\langle b\rangle_i$. Then both parties reconstruct $e$ and $f$ and $P_i$ lets $\langle uv\rangle_i=i\cdot ef + \langle a\rangle_i f + e\langle b\rangle_i + \langle c\rangle_i$. 
\par
\textbf{Oblivious Transfer and Multiplexer}. In 1-out-of-2 OT, a sender inputs two messages $(m_0,m_1)$ and a receiver inputs a choice bit $c \in \{0,1\}$. At the end of the protocol, the receiver obtains $m_c$, while the sender learns nothing. OT is usually realized by building a few base OT instances and extending them to a large number of instances. We use Ferret\cite{yang2020ferret} for OT extension, which is one of the PCG-style OT (\cite{slientOT1},\cite{slientOT2}) works. Multiplexer (MUX)
takes as input arithmetic shares of $a$ and boolean shares of
choice bit $c$ from two parties, and returns shares of $a$ if $c=1$, otherwise returns shares of 0. $\mathcal{F}_\text{MUX}$ can easily be implemented by two simultaneous calls to OT  \cite{CrypTFlow2}.
\par
\textbf{Homomorphic Encryption}. A homomorphic encryption of $x$
allows computing encryption of $f(x)$ without the knowledge of the
decryption key. In this work, we retain the matrix multiplication (Matmul) in Iron\cite{Iron}, which is based on the HE scheme of BFV (\cite{bfv1,bfv2}).

\section{Overview\label{chap:3}}
\subsection{Transformer Architecture}
\begin{figure}[htp]
\centerline{\includegraphics[width=.8\linewidth]{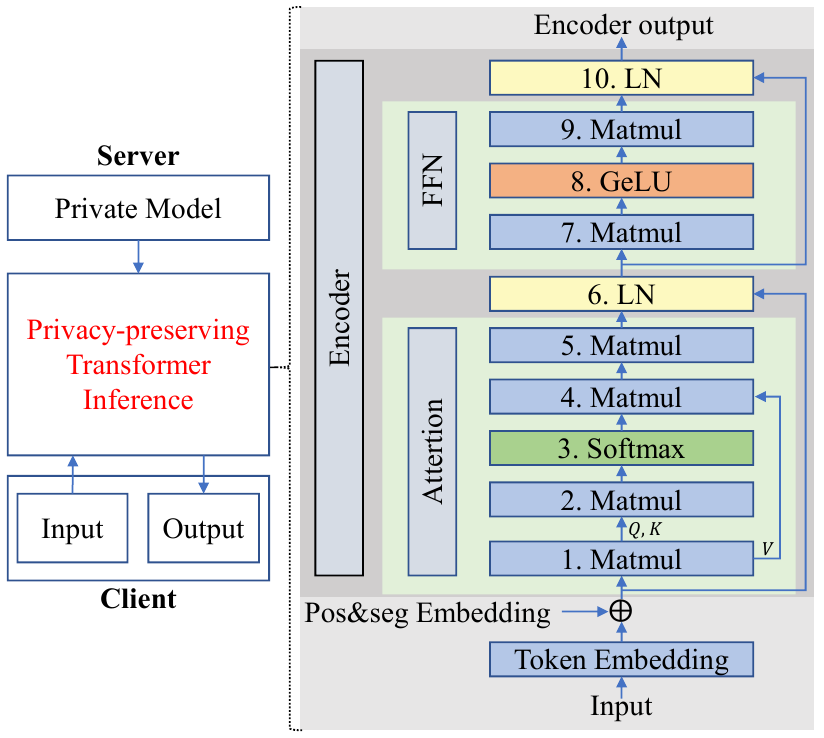}}
\caption{Overview of \emph{East}.}
\label{fig0}
\end{figure}
\par
\par
The encoder-decoder is used in Transformer. Because decoder is similar to encoder, we mainly focus on encoder in this paper.  The encoder contains several parts: the attention part, the feed-forward network (FFN) part, and two layer normalization (LN) parts, as shown in Fig. \ref{fig0}. The specific descriptions of these parts are as follows.

\textbf{Attention}. In the atteition part, the queries, keys and values are represented by matrices $Q,K$ and $V$. Let $d_k$ be the dimension of representations. The matrix of outputs is computed by
\begin{equation}
\begin{split}
\text{Attention}(Q, K, V) = \text{softmax}\left(\frac{QK^T}{\sqrt{d_k}}\right)V,
\end{split}
\end{equation}

\par
where softmax is computed by
\begin{equation}
\begin{split}
& \text{softmax}\left(x_0,x_1,\dots,x_{n-1}\right) \\
& =\frac{1}{\sum_{i=0}^{n-1}e^{x_i}}\left(e^{x_0},e^{x_1},\dots,e^{x_{n-1}}\right).
\end{split}
\end{equation}
\par
Multi-head attention extends the above mechanism to $H$ parallel attention layers, which can be represented as follows:
\begin{equation}
\begin{split}
\text{MultiHead} = \text{Concat}(\text{Attention}(Q_i, K_i, V_i),i\in [[H]])W^O.
\end{split}
\end{equation}
\par
\textbf{FFN}. FFN
consists of two linear layers with a GELU activation in between. The output can be obtained by
\begin{equation}
\begin{split}
\text{FFN}(X) = \text{GELU}(XW_{\text{FFN}_1} + B_{\text{FFN}_1})W_{\text{FFN}_2} + B_{\text{FFN}_2}.
\end{split}
\end{equation}
\par
\textbf{LN}. LN needs to compute the mean and variance of the input. For the input $\left(x_0,x_1,\dots,x_{n-1}\right)$, compute
\begin{equation}
\begin{split}
y_j & =\gamma_j\cdot\frac{x_j-\mu}{\sqrt{\sigma^2+\epsilon}}+\beta_j, \\
\text{where } \mu &  =\frac{1}{n}\sum_{j=0}^{n-1}x_j,\sigma^2=\frac{1}{n}\sum_{j=0}^{n-1}\left(x_j-\mu\right)^2.
\end{split}
\end{equation}
\par

\subsection{Privacy-preserving Transformer Inference}
In this paper, the encoder is divided into 10 layers. The specific computation tasks of each layer are shown in Table \ref{table1}. Layers 1-5 constitute the attention part. Layers 1-4 constitute the self-attention part and layer 5 is the output layer of multi-head attention. FFN consists of layers 7-9. Layers 6 and 10 are LN layers (including the addition of residuals). The division of $\sqrt{d_k}$ can be merged into the computation of $K$ or could be offset when computing truncation, such as adding another 3 bits for truncation when $\sqrt{d_k}$ is 8. According to our target model BERT, the module before encoders, such as the embedding layer, can be seen as a $\mathcal{F}_\text{Matmul}$ followed by a $\mathcal{F}_\text{LN}$. And the modules after encoders, such as the pooler layer and the fully connect layer, usually have simple architectures and can be seen as an additional FFN with the activation function using tanh.

\begin{table}[htb]
\centering
\caption{Hierarchical decomposition of the encoder.}
\linespread{1.3}\selectfont
\begin{tabular}{c|c|c}
\hline
No.  & Layer & Computation Task                                                                                                        \\ \hline
1 &
  Matmul &
  \begin{tabular}[c]{@{}c@{}}$Q=XW_Q+B_Q$\\ $K=XW_K+B_K$\\ $V=XW_V+B_V$\end{tabular} \\ \hline
2 &                   
  Matmul &
  $QK^T$ \\ \hline
3 &
  Softmax &
  $\text{softmax}\left(\frac{QK^T}{\sqrt{d_k}}\right)$ \\ \hline
4 &
  Matmul &
  $O_4=\text{softmax}\left(\frac{QK^T}{\sqrt{d_k}}\right)V$ \\ \hline
5 &
  Matmul &
  $O_5=O_4W_O+B_O$ \\ \hline
6 &
  LN &
  $O_6=\text{LN}(X+O_5)$ \\ \hline
7 &
  Matmul &
  $O_7=O_6W_{\text{FFN}_1}+B_{\text{FFN}_1}$ \\ \hline
8 &
  GELU &
  $O_8=\text{GELU}(O_7)$ \\ \hline
9 &
  Matmul &
  $O_9=O_8W_{\text{FFN}_2}+B_{\text{FFN}_2}$ \\ \hline
10 &
  LN &
  $X=O_{10}=\text{LN}(O_6+O_9)$ \\ \hline
\end{tabular}
\label{table1}
\end{table}

The above layers can be classified into linear layers and non-linear ones based on the underlying cryptographic protocols in our work.

\par
\textbf{Linear Layers}. Linear layers rely crucially on the matrix multiplication protocol $\mathcal{F}_\text{Matmul}$.  In the ciphertext-plaintext matrix multiplication, the server has weight $W$ and bias $B$ and the client has input data $X$. The client encrypts $X$ and sends $[X]$ to the server, and the server homomorphically computes $[Y]=[X]W+B$. 
The input and output can be shared between the two parties, and the bias can be added locally to the shares of the server. When computing $AB=(A_0+A_1)(B_0+B_1)=A_0B_0+A_0B_1+A_1B_0+A_1B_1$, where party $P_i, i=0,1$ holds $A_i, B_i$ such that $A=A_0+A_1, B=B_0+B_1$, $A_0B_0$ and $A_1B_1$ can be locally computed.  $A_0B_1$ and $A_1B_0$ can be obtained by two calls of ciphertext-plaintext matrix multiplications. Thus two calls of ciphertext-plaintext matrix multiplications could construct the ciphertext-ciphertext matrix multiplication. Iron\cite{Iron} gives efficient $\mathcal{F}_\text{Matmul}$ and we follow their work.
\par
\textbf{Non-linear Layers}. Non-linear layers, including $\mathcal{F}_\text{GELU}$, $\mathcal{F}_\text{Softmax}$ and $\mathcal{F}_\text{LN}$, have a relatively high proportion of overhead. We achieve high-quality polynomial approximation of GELU by using the NFGen \cite{NFGen} paradigm in plaintext, and propose a new OPPE algorithm for the oblivious evaluation of approximating polynomials, which can also be applied to other non-linear functions, e.g., tanh and exponent. More details of our OPPE and activation functions can be found in Section \ref{chap:4.1}. We design privacy-preserving protocols for softmax and LN with detailed consideration. More precisely, we make some conversions for effective computation of softmax and LN. The reciprocal and square-root reciprocal are implemented by Newton's method. The important point of Newton's method is to confirm the initial value and the number of iterations, which can be automatically determined by our error-limited determination method. $\mathcal{F}_\text{Softmax}$ and $\mathcal{F}_\text{LN}$ will be introduced in Section \ref{chap:4.2}.
\par
Then privacy-preserving Transformer can be built layer by layer with the operators mentioned above for inference. Further optimizations in detail will be conducted in Section \ref{chap:4.3} to improve the overall performance. 

\section{Supporting Protocols\label{chap:4}}
\subsection{Protocols for OPPE and Activation Functions\label{chap:4.1}}
\textbf{Oblivious Piecewise Polynomial Evaluation (OPPE)}. We briefly recall the high-level intuition of the OPPE algorithm from NFGen \cite{NFGen}.
For $m$ pieces with degree $d$ polynomials $p_m^d$ in $(w_0,w_m]$, let $\bm{w}=\left(w_0,w_1,\dots,w_{m-1}\right)$ be the break points (without $w_{m}$) and $I=\left(I_0,I_1,\dots,I_{m-1}\right), I_i=\left(w_i,w_{i+1}\right],i\in[[m]]$. Let $C=\left(\bm{c}_0,\bm{c}_1,\dots,\bm{c}_{m-1}\right)$ and $\bm{c}_i=\left(c_{i,0},c_{i,1},\dots,c_{i,d}\right)$ be the coefficients for $I_i$. For an input $x$, NFGen \cite{NFGen} first compares $x$ with each break point in $\bm{w}$ to get $\bm{comp}=\left(1\{x>w_0\},1\{x>w_1\},\dots,1\{x>w_{m-1}\}\right)$, and subtracts the results of adjacent intervals to get $\bm{mask}=\bm{comp} - \text{leftshift}(\bm{comp}, 1)$. Thus, the coefficients according to $x$ can be computed as $coeff_j = \sum_{i=0}^{m-1}mask_i\cdot c_{i,j}, j \in [[d]]$ and the final result can be obtained by computing $y=\sum_{j=0}^{d}coeff_j\cdot {xitem}_j$, where ${\bm{xitem}}=(1,x,\dots,x^d)$. 
\par
We notice that $\bm{mask}$ is obtained by a subtraction on $\bm{comp}$ and is used in a multiplication to compute $\bm{coeff}$.  The comparison result is usually a BShare, but subtraction and multiplication require the form of an AShare. Intuitively, a B2A is used to convert from BShare to AShare. We present our new OPPE $\mathcal{F}_\text{OPPE}$ in Alg. \ref{alg1}, which computes the result in a new way without B2A. An extra fruit of our OPPE is that the multiplication can be replaced by a multiplexer. The multiplexer has lower communication compared to the multiplication using Beaver triples. 
\par
\begin{algorithm}[htb]
        \caption{OPPE}
        \textbf{Public Parameters}: $p_m^d$.
        \begin{algorithmic}[1] 
        \Require $\langle x \rangle$.
        \Ensure $\langle y \rangle$, $y = p_m^d(x)$.
        \State $\langle \bm{comp}\rangle \leftarrow \text{GT}\left(\langle x\rangle,\bm{w}\right)$.
       
        \State $\langle \bm{xitem}\rangle \leftarrow \text{CalculateKx}\left(\langle x \rangle, d\right)$
        \State $\langle y_i\rangle \leftarrow \sum_{j=0}^{d}c_{i,j}\cdot \langle {xitem}_j\rangle$.
        \State $\langle z_0\rangle \leftarrow \langle y_0\rangle, \langle z_i\rangle \leftarrow \langle y_{i}\rangle - \langle y_{i-1}\rangle,1 \le i\le m-1$.
        \State $\langle \bm{v}\rangle \leftarrow \text{MUX}(\langle \bm{comp}\rangle, \langle \bm{z}\rangle)$.
        
        \State $\langle y\rangle \leftarrow \sum_{i=0}^{m-1}\langle v_i \rangle$.
        \State \Return $\langle y\rangle.$
        \end{algorithmic}
        \textbf{Function} $\text{CalculateKx}\left(\langle x \rangle,d\right):$
        \begin{algorithmic}[1] 
        \State $j \leftarrow 1.$
        \While {$j\le d$}
        	\State $t \leftarrow \min(j,t \leftarrow d-j).$
        	\State $(\langle x^{j+1}\rangle, ...,\langle x^{j+t}\rangle) \leftarrow \text{MUL}((\langle x^{1}\rangle, ...,\langle x^{t}\rangle),\langle x^{j}\rangle).$
	        \State $j \leftarrow 2j.$
        \EndWhile
        \State \Return $(\langle 1\rangle, \langle x^1\rangle,\dots,\langle x^{d}\rangle).$
        \end{algorithmic}
         
        \label{alg1}
    \end{algorithm}
    
\par
The core idea of our OPPE is to compute the result of each polynomial $y_i$ for each $I_i$, and the final result can be obtained by $y=comp_0\cdot y_0 + \sum_{i=1}^{m-1}{comp_i\cdot (y_{i}-y_{i-1})}$. Secure comparison is used to compute $\bm{comp}$ and  $\lfloor \log d\rfloor$ rounds of element-wise ciphertext-ciphertext multiplication to obtain the powers of $x$. 
Two parties can compute $\langle y_{i}\rangle$ locally, and then the multiplication of $\bm{comp}$ could be done by $\mathcal{F}_\text{MUX}$. Thus, our OPPE avoids B2A and further reduces the cost of multiplication by $\mathcal{F}_\text{MUX}$. Our idea can be extended to 3PC case, which makes piecewise polynomial evaluation more efficient than the one in ABY$^3$ since it computes the result by $y=\sum_{i=0}^{m-1}{(comp_i\wedge \neg comp_{i+1})\cdot y_i}$ and requires additional AND operations on the comparison results.
\par
OPPE could help to get the result of complex non-linear functions in the given finite range, but it is hard to tackle the case with an infinite range. Fortunately, many activation functions have asymptotic properties when the input tends to infinity, such as GELU and tanh. Now we apply our OPPE to the activation functions GELU and tanh. 
\par
\textbf{GELU}.
Notice that if $x \rightarrow - \infty, \text{GELU}(x) \rightarrow 0$, and if $x \rightarrow + \infty, \text{GELU}(x) \rightarrow x$. OPPE will return 0 if $x$ is less than or equal to the first break point, and we can set the polynomial of the rightmost interval to $f(x)=x$ and note the rightmost interval as $(w_{m-1},+\infty)$. Thus, the error can be limited to an acceptable range by selecting the appropriate $w_0$ and $w_{m-1}$. 
\par
\textbf{Tanh}.
Notice that if $x \rightarrow - \infty, \text{tanh}(x) \rightarrow -1$, and if $x \rightarrow + \infty, \text{tanh}(x) \rightarrow 1$. We shift $\text{tanh}(x)$ up vertically by one unit, denoted by $\text{tanh1}(x)=\text{tanh}(x)+1$. Notice that if $x \rightarrow - \infty, \text{tanh1}(x) \rightarrow 0$, and if $x \rightarrow + \infty, \text{tanh1}(x) \rightarrow 2$. Similar to GELU, set the polynomial of the rightmost interval to $f(x)=2$. Then $\text{tanh1}(x)$ can be computed by OPPE and $\text{tanh}(x)$ can be obtained by $\text{tanh}(x)=\text{tanh1}(x)-1$.
\subsection{Protocols for Softmax and LN\label{chap:4.2}}

\textbf{Softmax}.
There are two main difficulties in the computation of softmax: computing the exponent and computing the reciprocal. When the input is large, the result of the exponent will exceed the limit of FXP. 
The other difficulty is to compute the reciprocal of secret values. Thanks to Newton's method, the approximate value of the reciprocal can be obtained through multiple iterations. However, how to select the initial value and the number of iterations within the given error is something worth considering.
\par
Notice that there is an equivalence to computing softmax
\begin{equation}
\begin{split}
& \text{softmax}\left(x_0,x_1,\dots,x_{n-1}\right) \\
& =\text{softmax}\left(x_0-\tau,x_1-\tau,\dots,x_{n-1}-\tau\right)\\
& =\frac{1}{\sum_{i=0}^{n-1}e^{x_{i}-\tau}}\left(e^{x_0-\tau},e^{x_1-\tau},\dots,e^{x_{n-1}-\tau}\right).
\end{split}
\end{equation}
\par
Let $\bm{z}=\left(x_0-\tau,x_1-\tau,\dots,x_{n-1}-\tau \right)$ where $\tau=\max\left(x_0,\dots,x_{n-1}\right)$. Then the input of the exponent is less than or equal to 0, and the output of the exponent is in $(0,1]$, avoiding the overflow. And notice that if $x \rightarrow - \infty, e^x \rightarrow 0$, thus $e^x(x\le 0)$ can be computed by our OPPE in Section \ref{chap:4.1}. 
\par
$\sum_{i=0}^{n-1}e^{x_{i}-\tau}$ can be obtained by local addition with shares. And then the reciprocal can be computed by Newton's method. Alg. \ref{alg2} shows the protocol of reciprocal $\mathcal{F}_\text{Recip}$, which can obtain the approximate reciprocal within limited error. Our proposed determination method is executed in plaintext, and then the reciprocal is computed by Newton's method. 

\par
\begin{algorithm}[htb]
        \caption{Reciprocal}
        \textbf{Public Parameters}: $y_0,t\leftarrow\text{Determine}(a,b,\delta)$.
        \begin{algorithmic}[1] 
        \Require $\langle x \rangle$.
        \Ensure $\langle y \rangle$, $y \approx \frac{1}{x}$.
        \For {$i$ from 0 to $t-1$}
        \State $\langle y_{i+1} \rangle\leftarrow \text{MUL}(\langle y_{i} \rangle,2-\text{MUL}(\langle x \rangle, \langle y_{i} \rangle))$.
        \EndFor
        \State $\langle y\rangle\leftarrow\langle y_{t} \rangle.$
        \State \Return $\langle y\rangle.$
        \end{algorithmic}
        \textbf{Function} $\text{Determine}(a, b, \delta):$
        \begin{algorithmic}[1] 
        \State $L_{a,0}\leftarrow \frac{1}{a}-\delta.$
        \State $L_{b,0}\leftarrow \frac{1}{b}-\delta.$
        \State $H_{b,0}\leftarrow \frac{1}{b}+\delta.$
        \State $i \leftarrow 0.$
        \While {$H_{b,i}< L_{a,i}$}
        	\State solve $y\left(2-ay\right)\geq L_{a,i}$ to obtain $ [L_{a,i+1}, H_{a,i+1}].$
        	\State solve $y\left(2-by\right)\geq L_{b,i}$ to obtain $ [L_{b,i+1}, H_{b,i+1}].$
	        \State $i \leftarrow i+1.$
        \EndWhile
        \State $y_0 \leftarrow \frac{H_{b,i}+ L_{a,i}}{2}.$
        \State $t \leftarrow i.$
        
        \State \Return $y_0,t.$
        \end{algorithmic}
        \label{alg2}
    \end{algorithm}

\par 
\begin{algorithm}[htb]
        \caption{Softmax}
        \begin{algorithmic}[1] 
        \Require $\langle \bm{x}\rangle = (\langle x_{0} \rangle,\langle x_{1} \rangle,\dots,\langle x_{n-1} \rangle)$.
        \Ensure $\langle \bm{y}\rangle = (\langle y_{0} \rangle,\langle y_{1} \rangle,\dots,\langle y_{n-1} \rangle)$, $\bm{y} \approx \text{softmax}(\bm{x})$.
        
        \State $\langle \tau \rangle\leftarrow \text{MAX}(\langle x_{0} \rangle,\langle x_{1} \rangle,\dots,\langle x_{n-1} \rangle)$.
        \For {$i$ from 0 to $n-1$}
        \State $\langle z_{i} \rangle\leftarrow \langle x_{i} \rangle-\langle \tau \rangle$.
        \EndFor
        \State $\langle \bm{exp} \rangle \leftarrow \text{OPPE}_{exp}(\langle \bm{z} \rangle).$
        \State $\langle s \rangle \leftarrow \sum_{i=0}^{n-1}\langle exp_i \rangle.$
        \State $\langle \frac{1}{s} \rangle \leftarrow \text{Reciprocal}(\langle s \rangle).$
        \State $\langle \bm{y} \rangle\leftarrow \text{MUL}(\bm{exp}, \langle \frac{1}{s} \rangle)$.
        \State \Return $\langle \bm{y}\rangle.$
        \end{algorithmic}
        \label{alg3}
    \end{algorithm}

\par
Next, we will explain the idea behind our determination method. After each iteration, the error between the approximate value and the target value will gradually decrease until it reaches 0.  An interesting fact is that the error range of the previous iteration can be computed in reverse based on the given error of the current iteration. Given a small error, the result is within a small range near the target value. This range can be expanded by using reverse iterations, then limiting the final result to a small range is equivalent to limiting the value of the reverse iteration to a larger range. We perform reverse iterations on the left and right endpoints of the interval separately. When their ranges intersect after reverse iterations, the elements in the intersection can be selected as the initial values, because the error between the target value and the result obtained through the corresponding number of iterations does not exceed the given error. 
\par
For reciprocal, given interval $[a,b](a,b>0)$ and limited error $\delta(\delta>0)$, the reciprocal is in $[\frac{1}{b},\frac{1}{a}]$. The initial value is chosen from $[\frac{1}{b},\frac{1}{a}]$ and the intermediate result is in $(0,\frac{1}{a}]$. 
In Newton's method, for the reciprocal of $x$, the computation in each iteration is $y_{i+1}=y_i\left(2-xy_i\right)$. 
The initial value and iteration number are determined based on the given error. We iteratively compute $[L_{a},H_{a}]$ and $[L_{b},H_{b}]$, which requires solving the quadratic inequalities. When $H_{b}\ge L_{a}$, any value in $[L_{a},H_{b}]$ can meet the requirements for limited error. We determine the minimum number of iterations through successive iterations and set the initial value to the average of final $H_{b}$ and $L_{a}$. The number of iterations  is set to  the number of reverse iterations.
\par
Now, secure protocol for softmax can be presented using the above methods, as shown in Alg \ref{alg3}. We use $\mathcal{F}_\text{MAX}$ to compute the maximum of the input, which can be instantiated by Maxpool$_d$ protocol \cite{CrypTFlow2}. Then use our $\mathcal{F}_\text{OPPE}$ and $\mathcal{F}_\text{Recip}$ to obtain the result of the exponent and reciprocal. Notice that for the input with length $n$, the input for $\mathcal{F}_\text{Recip}$ is in $[1,n]$ since the maximum element of the exponent is 0 and all other elements are less than 0. 
Finally, a $\mathcal{F}_\text{MUL}$ is used to obtain the result of softmax. 

\par
\textbf{LN}.
Computing the mean and variance on secret values needs secure division. We first perform an equivalent transformation on LN to eliminate the division. For input $\left(x_0,x_1,\dots,x_{n-1}\right)$, the original computation in LN is 
\begin{equation}
\begin{split}
y_j & =\gamma_j\cdot\frac{x_j-\mu}{\sqrt{\sigma^2+\epsilon}}+\beta_j.
\end{split}
\end{equation}
\par
That is equal to 
\begin{equation}
\begin{split}
y_j 
& =\gamma_j\cdot\frac{n(x_j-\mu)}{n\sqrt{\sigma^2+\epsilon}}+\beta_j \\
& =\gamma_j\cdot\frac{n\left(x_j-\mu\right)\ }{n\sqrt{\frac{1}{n}\sum_{j=0}^{n-1}\left(x_j-\mu\right)^2+\epsilon}}+\beta_j \\
& =\sqrt n\gamma_j\cdot\frac{nx_j-n\mu}{\sqrt{\sum_{j=0}^{n-1}\left(nx_j-n\mu\right)^2+\epsilon'}}+\beta_j.
\end{split}
\end{equation}
\par
Let $z_j=nx_j-n\mu=nx_j-\sum_{j=0}^{n-1}x_j$, then 
\begin{equation}
\begin{split}
y_j =\sqrt n\gamma_j\cdot\frac{1}{\sqrt{\sum_{j=0}^{n-1}z_j^2+\epsilon'}}\cdot z_j+\beta_j.
\end{split}
\end{equation}
\par
Secure protocol for LN is presented in Alg \ref{alg4}. The server multiplies its own $\bm{\gamma}$ by $\sqrt n$. Then the conversion is done locally by the two parties, and the variance is obtained by $\mathcal{F}_\text{Matmul}$. $\mathcal{F}_\text{InvSqrt}$ refers to the protocol of square-root reciprocal,  which is similar to $\mathcal{F}_\text{Recip}$, except that the computation in each iteration becomes $y_{i+1}=\frac{1}{2}y_i\left(3-xy_i^2\right)$ and the quadratic inequalities become cubic inequalities. The  result of normalization is obtained by $\mathcal{F}_\text{MUL}$ and then sent to a linear operation to get the final result. Elements in different positions can be seen as elements from different channels.
At this point, the linear operation of LN is equivalent to the linear operation of BN. In addition, when the variance of the input is large, we can proportionally reduce the input to prevent overflow of intermediate results according to the normalization property.
\par 
\begin{algorithm}[htb]
        \caption{LN}
        \begin{algorithmic}[1] 
        \Require $\langle \bm{x}\rangle = (\langle x_{0} \rangle,\langle x_{1} \rangle,\dots,\langle x_{n-1} \rangle)$, $\bm{\gamma}$ and $\bm{\beta}$ from the server in addition.
        \Ensure $\langle \bm{y}\rangle = (\langle y_{0} \rangle,\langle y_{1} \rangle,\dots,\langle y_{n-1} \rangle)$, $\bm{y} \approx \text{LN}_{\bm{\gamma}, \bm{\beta}}(\bm{x})$.
        \State server: $\bm{\gamma} \leftarrow \sqrt{n}\bm{\gamma}.$
        \State $\langle n\mu \rangle \leftarrow \sum_{j=0}^{n-1}\langle x_j \rangle.$
        \For {$i$ from 0 to $n-1$}
        \State $\langle z_{i} \rangle\leftarrow n\langle x_{i} \rangle-\langle n\mu \rangle$.
        \EndFor
        \State $\langle s \rangle\leftarrow \text{MUL}(\langle \bm{z} \rangle,\langle \bm{z} \rangle)+ \epsilon$.
        \State $\langle \eta \rangle\leftarrow \text{InvSqrt}(\langle s \rangle)$.
        \State $\langle norm \rangle\leftarrow \text{MUL}(\bm{z}, \langle \eta \rangle)$.
        \State $\langle \bm{y} \rangle\leftarrow \text{Linear}_{\bm{\gamma}, \bm{\beta}}(\langle norm \rangle)$.
        \State \Return $\langle \bm{y}\rangle.$
        \end{algorithmic}
        \label{alg4}
    \end{algorithm}
   
\par
\subsection{Further Optimizations\label{chap:4.3}}

\textbf{Combination of Linear Layers}.
The linear computation of $Q,K$ and $V$ in the attention part can be merged. That is 
\par
\begin{equation}
\begin{split}
Q =XW^Q+B^Q, K =XW^K+B^K,V =XW^V+B^V\\
\iff (Q,K,V)=X(W^Q,W^K,W^V)+(B^Q,B^K,B^V).
\end{split}
\end{equation}
\par
We have an improvement by computing $Q,K$ and $V$ in this way, compared to the independent computation of the three.
\par
\textbf{Truncation Optimization}.
Cheetah \cite{Cheetah} provides a protocol for truncation with known MSB that is more efficient than the original truncation. In fact, for a floating-point number $x$,  selecting the appropriate $M$ that satisfies $\left|x \right| \le M$, $\text{MSB}(\text{FXP}_{\ell,f}(x)+ \text{FXP}_{\ell,f} (M))$ is always 0, because $x+M\ge-|x|+M\ge 0$. Truncation for $\text{FXP}_{\ell,f}(x)+ \text{FXP}_{\ell,f} (M)$ allows for the use of truncation with known MSB. Then, the truncation for $\text{FXP}_{\ell,f}(x)$ can be obtained by subtracting $\text{FXP}_{\ell,f}(M)/2^f$ from the truncation for $\text{FXP}_{\ell,f}(x)+ \text{FXP}_{\ell,f} (M)$. When $M$ is greater than the absolute values of all the intermediate results, the original truncation can be replaced by the truncation with known MSB.
\par
\textbf{Processing of Packed Communication}.
When performing the same computation on multiple sets of data, it is better to execute simultaneously, which will save the overall costs, especially the number of communication rounds. 
\par
First, element-wise computation can be processed simultaneously, whether the input is in vector form or matrix form, such as activation functions and element-wise multiplication. Then, the computation between different rows in a matrix can be processed simultaneously, such as the computation of the reciprocal in softmax and the square-root reciprocal  in LN, and this  could also be applied to  multiple matrix forms in multi-head attention. Finally, for LN, the elements between different columns can be considered as elements of different channels, so that the final linear transformation is equivalent to the  linear transformation of BN.

\par
\textbf{Handling of Padding and Masking}.
Due to the different number of tokens of the input, the approach of ``padding then masking" is usually used to obtain a fixed-length input in plaintext inference. Many of the intermediate results are meaningless in this case. We could compute only the corresponding part of the actual processing length, which significantly reduces costs for a short input.
\par
\subsection{Security Analysis\label{chap:4.4}}
 Similar to the security of protocols in \cite{CrypTFlow2},\cite{Iron}, our protocols directly follow the hybrid
model. In particular, the security of the GELU is easy to see in (CMP, MUL, MUX)-hybrid. The security of the softmax protocol follows in
(MAX, OPPE, Recip, MUL)-hybrid. Besides, the security of the LN protocol follows in
(invSqrt, MUL, Linear)-hybrid.
\par
\section{Evaluations\label{chap:7}}

\textbf{Setup}. 
\emph{East} is built on top of the SEAL library \cite{seal}, the EMP toolkit \cite{emp-toolkit} and the EzPC-based frameworks \cite{ezpc},\cite{CrypTFlow2},\cite{Cheetah}. 
We simulate the same LAN network setting with Iron\cite{Iron}, where the bandwidth is 377 MBps and the echo latency is 0.8 ms.
All the following experiments are performed on Ubuntu 20.04 with an Intel(R) Core(TM) i9-10900X at 3.70 GHz. We run our benchmarks on BERT (BERT-Base) \cite{bert} and the dataset is the Stanford Sentiment Treebank (SST-2) from GLUE benchmarks \cite{glue}. Notice that the experiments of \emph{East} use a single-threaded setting, while a multi-threaded setting will obtain a better performance.
\par
\begin{figure}[htp]
\centerline{\includegraphics[width=\linewidth]{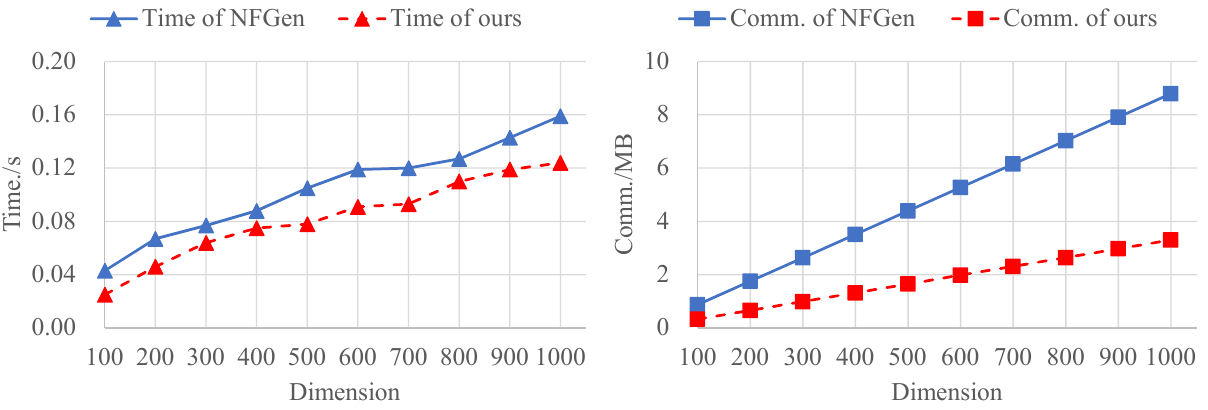}}
\caption{OPPE performance of NFGen and ours.}
\label{fig1}
\end{figure}

\textbf{Evaluation of OPPE and Activation Functions}. 
We compare our OPPE with the original scheme in NFGen \cite{NFGen} for different dimensions of input with parameters $m=8,d=3$. We implement the OPPE from NFGen \cite{NFGen} and ours in the same setting. Fig. \ref{fig1} shows the performance of the total for OPPE of NFGen \cite{NFGen} and ours. Both NFGen \cite{NFGen} and our work have linear growth with the input dimension in terms of runtime and communication, and our OPPE has a better performance.
\par
\par
We apply OPPE to exponent (Exp), GELU, and tanh, and the  runtime and communication are shown in Table \ref{table2}. We set $m=8,d=3$ for Exp and GELU, and $m=7,d=3$ for tanh. The dimensions of Exp, GELU, and tanh are 196608, 393216, and 768, respectively. 
\par
\begin{table}[htb]
\centering
\caption{Performance comparison of NFGen and ours. (Time is in seconds and communication is in MB.)}
\linespread{1.3}\selectfont
\setlength{\tabcolsep}{1mm}
\begin{tabular}{|c|c|cc|cc|cc|}
\hline
\multirow{2}{*}{Work} &
  \multirow{2}{*}{Function} &
  \multicolumn{2}{c|}{Offline} &
  \multicolumn{2}{c|}{Online} &
  \multicolumn{2}{c|}{Total} \\ \cline{3-8} 
 &
   &
  \multicolumn{1}{c|}{Time} &
  Comm. &
  \multicolumn{1}{c|}{Time} &
  Comm. &
  \multicolumn{1}{c|}{Time} &
  Comm. \\ \hline
\multirow{3}{*}{NFGen} &
  Exp &
  \multicolumn{1}{c|}{19.69} &
  1601.71 &
  \multicolumn{1}{c|}{23.18} &
  138.71 &
  \multicolumn{1}{c|}{42.87} &
  1740.42 \\ \cline{2-8} 
 &
  GELU &
  \multicolumn{1}{c|}{36.03} &
  3203.42 &
  \multicolumn{1}{c|}{45.97} &
  278.49 &
  \multicolumn{1}{c|}{82.00} &
  3481.91 \\ \cline{2-8} 
 &
  Tanh &
  \multicolumn{1}{c|}{0.06} &
  6.24 &
  \multicolumn{1}{c|}{0.07} &
  0.47 &
  \multicolumn{1}{c|}{0.13} &
  6.71 \\ \hline
\multirow{6}{*}{\textbf{Ours}} & \multirow{2}{*}{Exp} & \multicolumn{1}{c|}{5.10} & 533.19 & \multicolumn{1}{c|}{23.30} & 124.97 & \multicolumn{1}{c|}{28.41} & 658.16 \\ \cline{3-8} 
 &
   &
  \multicolumn{1}{c|}{(3.9$\times$)} &
  (3.0$\times$) &
  \multicolumn{1}{c|}{(1.0$\times$)} &
  (1.1$\times$) &
  \multicolumn{1}{c|}{(1.5$\times$)} &
  (2.6$\times$) \\ \cline{2-8} 
 &
  \multirow{2}{*}{GELU} &
  \multicolumn{1}{c|}{12.57} &
  1067.88 &
  \multicolumn{1}{c|}{47.45} &
  249.29 &
  \multicolumn{1}{c|}{60.02} &
  1317.17 \\ \cline{3-8} 
 &
   &
  \multicolumn{1}{c|}{(2.9$\times$)} &
  (3.0$\times$) &
  \multicolumn{1}{c|}{(1.0$\times$)} &
  (1.1$\times$) &
  \multicolumn{1}{c|}{(1.4$\times$)} &
  (2.6$\times$) \\ \cline{2-8} 
 &
  \multirow{2}{*}{Tanh} &
  \multicolumn{1}{c|}{0.02} &
  2.08 &
  \multicolumn{1}{c|}{0.07} &
  0.41 &
  \multicolumn{1}{c|}{0.08} &
  2.49 \\ \cline{3-8} 
 &
   &
  \multicolumn{1}{c|}{(3.3$\times$)} &
  (3.0$\times$) &
  \multicolumn{1}{c|}{(1.1$\times$)} &
  (1.2$\times$) &
  \multicolumn{1}{c|}{(1.5$\times$)} &
  (2.7$\times$) \\ \hline
\end{tabular}
\label{table2}
\end{table}
\par
During the online phase, our OPPE has a close runtime with NFGen \cite{NFGen} and achieves 1.1$\times$ lower communication. Moreover, due to the reduction in the use of Beaver triples, we obtain about 3$\times$ optimization during the offline phase. In total, we achieve about 1.5$\times$ lower runtime and 2.6$\times$ lower communication.
\par

\par
\begin{figure}[htp]
\centerline{\includegraphics[width=\linewidth]{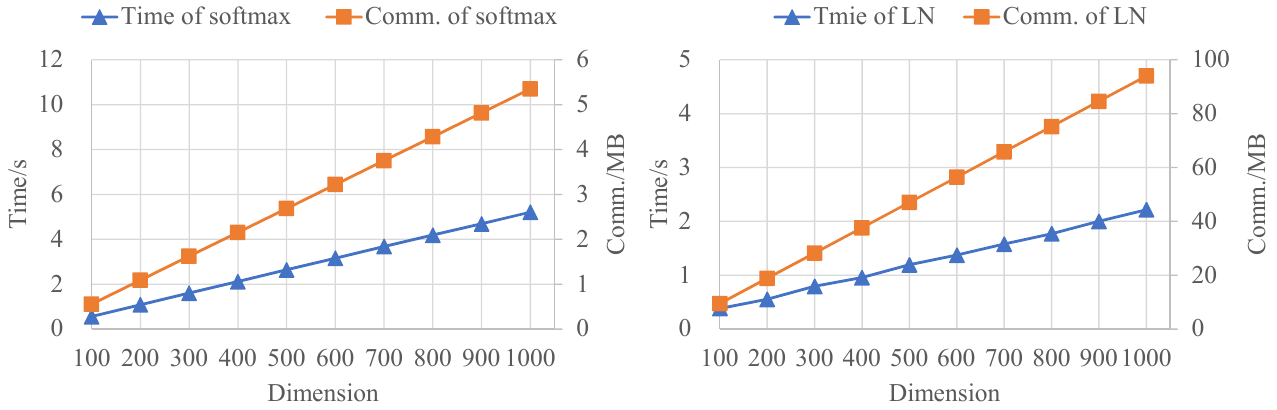}}
\caption{Performance of softmax and LN.}
\label{fig2}
\end{figure}

\textbf{Evaluation of Softmax and LN}. 
We show the performance of our softmax and LN protocols in Fig. \ref{fig2}. Both our softmax and LN have linear growth with the input dimension in terms of runtime and communication. When the input dimension is 1000, softmax requires approximately 5 seconds of runtime and 5 MB of communication, while LN requires approximately 2 seconds of runtime and less than 100 MB of communication.

\par

\textbf{Comparison with Iron}. 
We compare our work with Iron \cite{Iron} in Table \ref{table4}. The performance of Iron\cite{Iron} comes from Table 4 in Iron\cite{Iron}. For GELU, we achieve about 1.5$\times$ lower runtime and 2.9$\times$ lower communication, which shows that our OPPE is effective in processing the  activation functions. We achieve about 1.3$\times$ lower communication for softmax and  1.2$\times$ lower runtime for LN, compared with Iron \cite{Iron}. In total, we achieve about 1.2$\times$ lower runtime and 1.8$\times$ communication.  This means that for BERT-Base with 12 encoders, we will save about 256 seconds  of runtime and 30 GB of communication.

\par

\begin{table}[htb]
\centering
\caption{Performance comparison of Iron and ours. (Time is in seconds and communication is in MB.)}
\linespread{1.3}\selectfont
\setlength{\tabcolsep}{1mm}
\begin{tabular}{|c|c|c|c|c|c|}
\hline
Work & Metrics                    & GELU          & Softmax       & LN (2 calls)  & Total         \\ \hline
\multirow{2}{*}{Iron} & Time (Sec)                  & 88.08 & 30.31 & 13.05 & 131.44 \\ \cline{2-6} 
     & Comm.(MB)                  & 3848.30       & 1284.08       & 575.23        & 5707.61       \\ \hline
\multirow{4}{*}{\textbf{Ours}}                    & \multirow{2}{*}{Time (Sec)} & 60.02 & 39.42 & 10.63 & 110.07 \\ \cline{3-6} 
     &                            & (1.5$\times$) & (0.8$\times$) & (1.2$\times$) & (1.2$\times$) \\ \cline{2-6} 
     & \multirow{2}{*}{Comm.(MB)} & 1317.17       & 983.90        & 838.80        & 3139.87       \\ \cline{3-6} 
     &                            & (2.9$\times$) & (1.3$\times$) & (0.7$\times$) & (1.8$\times$) \\ \hline
\end{tabular}
\label{table4}
\end{table}
\par

\begin{figure}[htp]
\centerline{\includegraphics[width=.8\linewidth]{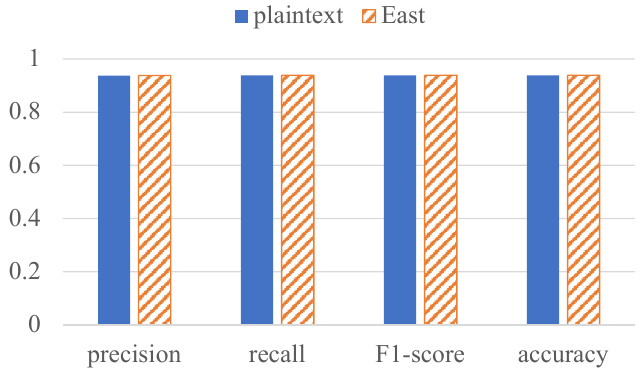}}
\caption{Inference comparison of plaintext and \emph{East}.}
\label{fig3}
\end{figure}

\textbf{Evaluation of Inference}. 
Fig. \ref{fig3} shows the comparison between our work and the plaintext  for four indicators, including precision, recall, F1-score, and accuracy. \emph{East} has limited the errors of the protocols of activation functions, reciprocal and square-root reciprocal, and achieves consistent results with plaintext inference.

\par
\par

\section{Conclusion\label{chap:8}}
In this paper, we presented \emph{East}, a privacy-preserving Transformer inference framework that faithfully maintains the architecture of the original model and protects the privacy of both client data and server model parameters. Specifically, we presented the protocols for secure evaluation of the activation functions used in Transformer, including GELU and tanh, by our new oblivious piecewise polynomial evaluation algorithm. Moreover, the protocols for softmax and layer normalization are carefully designed based on the proposed conversion methods and the error-limited determination methods. Further, we conducted several optimizations to enhance the overall performance. 
Experimental results showed that \emph{East} can achieve about 1.2$\times$ lower runtime and 1.8$\times$ communication than the state-of-the-art work and maintains inference accuracy with the plaintext model. We believe that our work will help promote the development of privacy-preserving Transformer inference.

\section*{Acknowledgment}
This paper is supported by the National Key R\&D Program of China (2021YFB2700200), and the National Natural Science Foundation of China (U21B2021, 61972018, 61932014, U2241213).
\bibliographystyle{unsrt}
\bibliography{refs}

\end{document}